\documentclass[
*aip,
reprint,
runinaddress,
showpacs,preprintnumbers,
 amsmath,amssymb,
 aps,
pra,
floatfix,
]{revtex4-1}
\usepackage{mathtools}
\usepackage{graphicx}% Include figure files
\usepackage{dcolumn}% Align table columns on decimal point
\usepackage{bm}% bold math
\usepackage{hhline}
\usepackage{color,soul}
\usepackage{float}
\usepackage{mathrsfs}
\usepackage{lipsum} % Generating random text 
\usepackage{romannum} % Roman numerals

\usepackage[normalem]{ulem}
\usepackage[dvipsnames]{xcolor}
\usepackage[sort&compress]{natbib}
\usepackage{xr-hyper}
\usepackage{hyperref}
\usepackage{float}
\hypersetup{
    colorlinks,
    linkcolor={blue},
    citecolor={blue},
    urlcolor={blue}
	}

\begin{document}
\pagenumbering{arabic} 
\title{A compact and highly collimated atomic/molecular beam source}

\author{Geetika Bhardwaj}%
\affiliation{Tata Institute of Fundamental Research Hyderabad, 36/P Gopanpally, Hyderabad 500046, Telangana, India}

\author{Saurabh Kumar Singh}%
\affiliation{Tata Institute of Fundamental Research Hyderabad, 36/P Gopanpally, Hyderabad 500046, Telangana, India}

\author{Pranav R. Shirhatti}
\email[Author to whom correspondence should be addressed. \\ e-mail:{\ }]{pranavrs@tifrh.res.in}
\affiliation{Tata Institute of Fundamental Research Hyderabad, 36/P Gopanpally, Hyderabad 500046, Telangana, India}%

\begin{abstract}
\textbf{Abstract:} 
We describe the design, characterization and application of a simple, highly collimated and compact atomic/molecular beam source.
This source is based on a segmented capillary design, constructed using a syringe needle. 
Angular width measurements and free molecular flow simulations show that the segmented structure effectively suppresses atoms travelling in off-axis directions, resulting in a narrow beam of Helium atoms having a width of 7 mrad (full width half maximum).
We demonstrate an application of this source by using it for  monitoring real-time changes in surface coverage on a clean Cu(110) surface exposed to oxygen, by measuring specular reflectivity of the Helium beam generated using this source. 
\end{abstract}

\maketitle

\section{\label{sec:level1}Introduction}

Atomic and molecular beam techniques find widespread use in areas ranging from fundamental scientific studies to important technological applications.  
They play a crucial role in high resolution atomic and molecular spectroscopy measurements, studies involving cold atoms, understanding energy transfer in intermolecular and molecule - surface collisions, dynamics of chemical reactions, surface chemistry, thin film and coating technology to name a few \cite{scoles_book, campargue_book_2001}.

In the context of understanding physical and chemical process on surfaces, atomic/molecular beam - surface scattering experiments are very valuable.
For example, techniques based on Helium atom scattering (HAS) from surfaces provide a wide range of information ranging from changes in surface coverage, phonon energy spectrum and dynamics, surface adsorbate motion, crystalline nature and even structural features by means of microscopy \cite{farias_rev1998, holst_perspective_2021}.
In particular, specular reflection of He atoms from surfaces is a highly sensitive technique to measure small changes in adsorbate coverage, especially on flat single crystal surfaces.
Here, diffuse scattering of incident He atoms caused by adsorbate induced disorder on the surface leads to a decrease in specular reflected He signal with increasing adsorbate coverage.
Typically, diffuse scattering cross sections of He from surface adsorbates are much larger than that expected from Van der Waals radii and are of the order of 100 \AA$^2$ per adsorbed molecule \cite{comsa_diffuse_1983, comsa_HeScat_rev_1985, holst_bracco_bookCh_2013}.
As a result, small changes in surface coverage of the order of 0.01 monolayer (ML) can be detected.
Further, use of thermal energy He beams typically having incidence kinetic energy $<$ 100 meV means that this technique is soft and non-destructive.
These features make specular He reflectivity an excellent tool for measuring sticking probabilities of adsorbates on surfaces.

An elegant strategy to measure surface coverage using He reflectivity was demonstrated by Higgins and co-workers \cite{scoles_CH4_diss_2001}. 
In their studies of quantum state resolved chemisorption of CH$_4$ on a Pt(111) surface
they used a seeded beam of CH$_4$ in He, at an incidence angle of 45$^\circ$.
Change in specular scattered He was used to estimate the surface coverage resulting from the dissociation of CH$_4$ and to evaluate the initial sticking probabilities.
It should be noted that a large incidence angle of 45$^\circ$ (needed for using He specular reflection as a probe) limits the kinetic energy associated with the normal component of incident momentum, thereby making the study of reactions with large incidence energy thresholds very difficult using this approach.
Using an independent He beam at large angle from surface normal (for increased diffuse scattering cross section) as a probe, with the incident molecular beam (reactant) near surface normal, can in principle circumvent this limitation. 
However, a typical design for producing a well-collimated He beam consists of a series (2-3) of differentially pumped vacuum chambers with a large footprint. 
This makes it very difficult to integrate a well-collimated He atom source with molecule-surface scattering experiments. 

Recent work by Li and co-workers \cite{raman_CascadedCollimator_2019}, where they demonstrate an extremely compact, on-chip collimated atomic beam source using segmented micro-channels etched on a silicon wafer, provides a route to overcome the above difficulty.
Building further on the ideas presented by Li and co-workers, we present a design of a simple, compact and highly collimated atom beam source which can be easily fabricated and incorporated into molecule - surface scattering experiments.
In our case, the atom beam source is based on a stainless steel capillary (a commercially available syringe needle), machined to have a segmented structure.
Further, we demonstrate an application of this compact and highly collimated atom beam source for measuring real-time surface coverage changes the case of dissociative chemisorption of oxygen on a clean Cu(110) surface, using specular He reflection as a probe.

\section{\label{sec:level1}Experimental setup}

\begin{figure*}[!]
	\centering
	\includegraphics[width = 1\linewidth, draft=false]{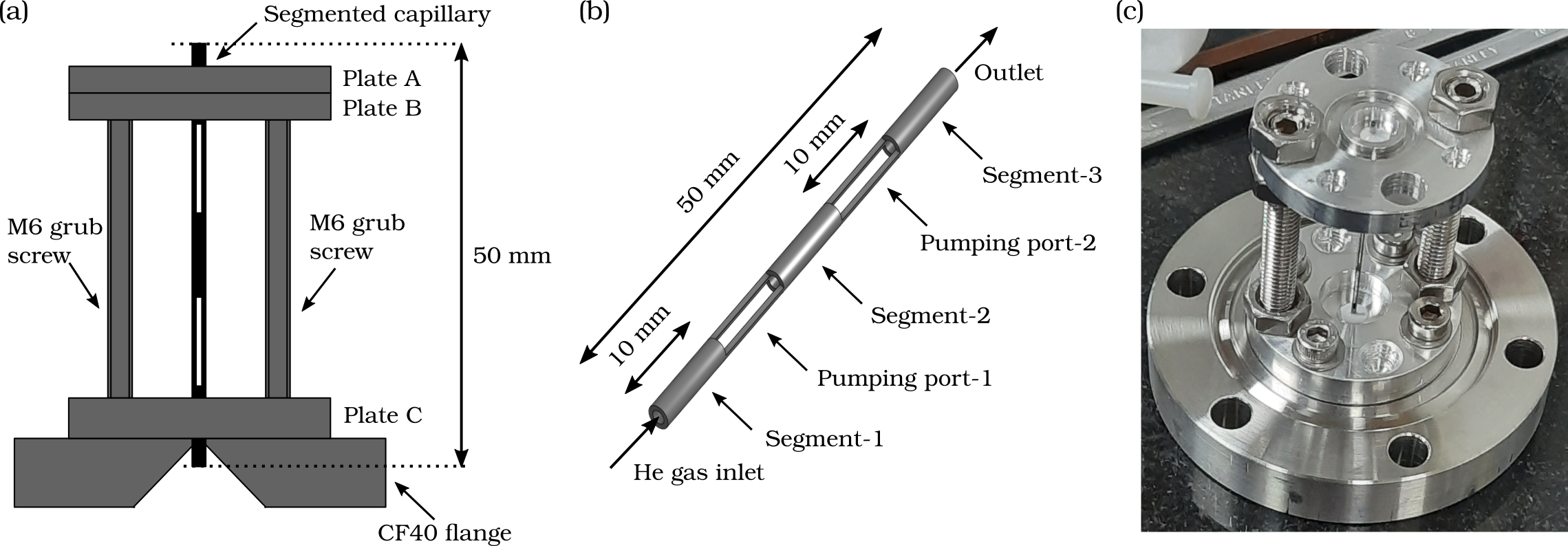}
	\caption{(a) Schematic diagram of the segmented capillary based atom beam source. 
	Plates A, B and C were used to hold the segmented needle and were secured using a pair of M6 grub screws.
	Entire assembly was mounted on a CF40 flange such that the He beam outlet was positioned at the center of a cone-shaped aperture on the flange.
    (b) A detailed view of the segmented capillary structure, made using a syringe needle. Its overall length was 50 mm, the inner diameter was 0.5 mm and the wall thickness was 0.25 mm. 
	Its walls were machined to create open regions of 10 mm length (pumping ports) between successive segments.
	He gas was leaked into the inlet side in a controlled manner using a needle valve and a collimated beam emerged from the outlet side.
	(c) Actual picture of our compact atom beam source.}
	\label{fig:collimator_panel}
\end{figure*}

\subsection{Segmented capillary based atomic beam source}

Figure \ref{fig:collimator_panel}a shows a schematic diagram of our compact atomic beam source.
A commercially available stainless steel capillary (syringe needle) with 0.5 mm inner diameter and 0.25 mm wall thickness, was machined using a hand-held grinding tool to make openings along its walls, resulting in a segmented structure with an overall length of 50 mm
(Fig. \ref{fig:collimator_panel}b).
The segmented capillary was mounted on a pair of metal plates (B and C) which were supported using two 6 mm (M6) grub screws mounted on a CF40 flange with a cone-shaped opening in its center. 
A leak tight seal among outer surface of the capillary and the supporting plates was achieved using vacuum compatible glue (Torr-Seal, two part epoxy sealant).
A picture of this assembly is shown in Fig. \ref{fig:collimator_panel}c.
The segmented capillary consists of three stages acting as long thin channels, each 10 mm in length (segments 1- 3, Fig. \ref{fig:collimator_panel}b). 
These are separated by two 10 mm long segments with openings along the walls on diametrically opposite sides, acting as pumping ports.
These openings allow the removal of atoms travelling in off-axis direction, eventually leading to a highly collimated beam emerging from the outlet.
Optimal dimensions of the capillary and individual segments were decided with the help of free molecular flow simulations carried out using the software Molflow+ \cite{molflow_web, molflow_kersevan_paper_2019}.

\subsection{Beam width characterization}

\begin{figure*}[t]
	\centering
	\includegraphics[width = 1\linewidth, draft=false]{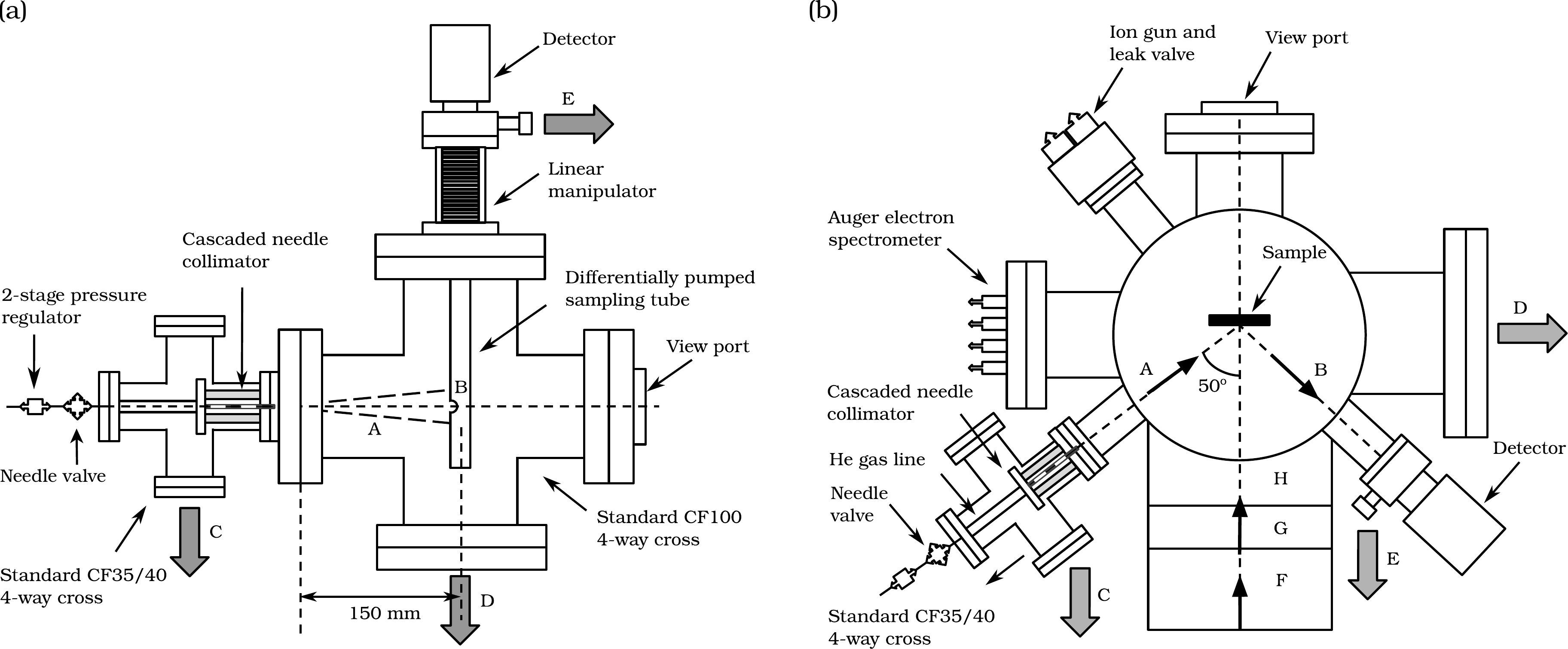}
	\caption{(a) Schematic diagram of the experimental setup to characterize the beam width. 
	The segmented capillary based source was placed in the CF40 four-way cross, which was attached to the detection chamber. 
	A differentially pumped sampling tube having a slit (1 mm width, 20 mm length) was moved in a plane perpendicular to the beam using a linear manipulator, for beam width measurement.
	Source to sampling plane distance was 160 mm.
	(b) Schematic diagram of the ultra high vacuum chamber setup used to measure the real-time surface coverage of Cu(110) exposed to oxygen, using specular reflection of He.
	Source chamber was attached at an angle of 50$^\circ$ (from target surface normal) and the source to target distance was 130 mm.
	Specular reflected flux of He from the Cu(110) surface was detected in a similar manner as in (a).
	This chamber is equipped with an ion source and Auger electron spectrometer for sample cleaning (sputtering) and chemical composition analysis, respectively.}
	\label{fig:expt_setup_panel}
\end{figure*}

Figure \ref{fig:expt_setup_panel}a shows a schematic diagram of the experimental setup used for measuring the width of the atomic beam generated using segmented capillary source.
This source was inserted in a CF40 4-way cross (source chamber), with a needle valve outside to control the He flow into the capillary. 
Source chamber was attached to a larger vacuum chamber, a  CF100 4-way cross (detection chamber). 
%%%%%%%%%%%%%%%%
Detector assembly consisted of a differentially pumped sampling tube (25.4 mm diameter) connected to a mass spectrometer (SRS RGA 200), with a slit shaped opening (approximately 1 mm width and 20 mm length).
Sampling tube and the mass spectrometer were mounted on a single-axis linear manipulator, allowing to move the detection assembly in a vertical plane (perpendicular to the atomic beam), thereby enabling angular width measurement.
Source chamber, detection chamber and the sampling tube were pumped using turbo molecular pumps (denoted by C, D and E) with the nominal speeds of 80 l/s (HiPace 80,
Pfeiffer Vacuum), 400 l/s (HiPace 400,
Pfeiffer Vacuum) and 80 l/s (HiPace 80,
Pfeiffer Vacuum), respectively. 
All the turbo pumps were backed by a single rotary vane pump with 11 m$^3$/hr pumping speed (Duo 11, Pfeiffer Vacuum). 
Typical steady state pressures (with beam off) in the source and detection chambers were 2$\times$10$^{-8}$ mbar and 1$\times$10$^{-8}$ mbar, respectively. 
To generate He atom beam, the needle valve was opened in a controlled manner to maintain a steady state pressure of around 3$\times$10$^{-5}$  mbar in the source chamber. 
The ultimate background partial pressure of He in the sampling tube was $\sim$ 3$\times$10$^{-11}$ mbar with the beam off and with the beam on, a maximum He signal (at the peak) of 3 $\times$10$^{-9}$ mbar was observed. 

\subsection{Surface coverage and sticking probability measurement using Helium reflectivity}

\begin{figure*}[t]
	\centering
	\includegraphics[width = 1\linewidth,trim={0cm 0cm 0cm 0cm}, draft=false]{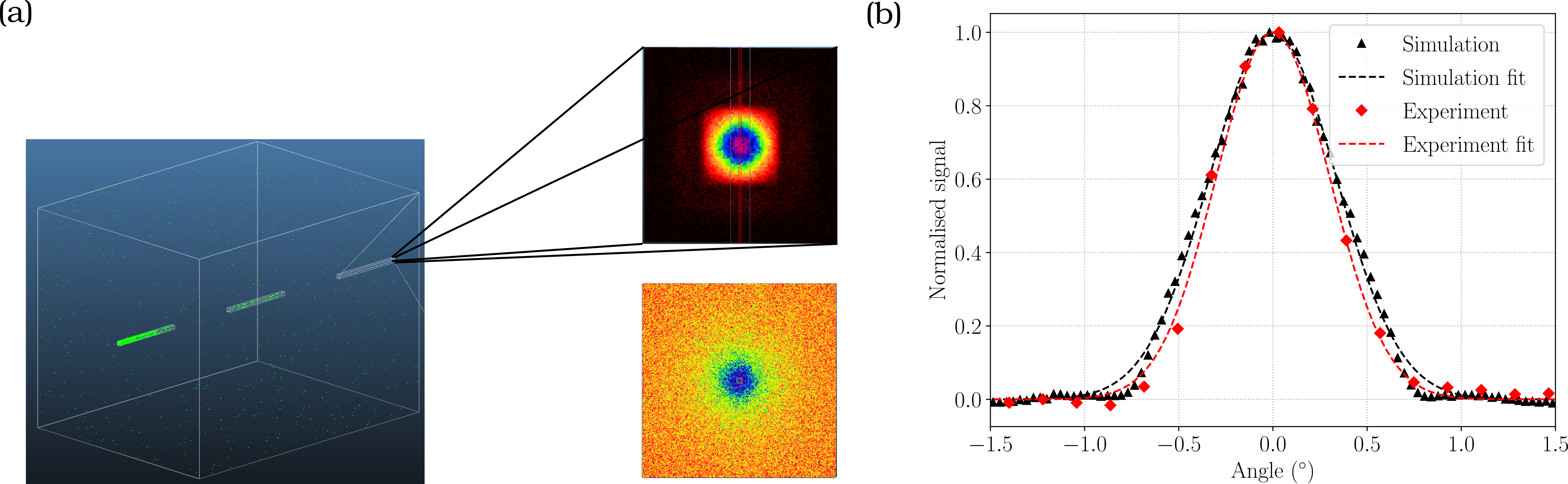}
	\caption{(a) Arrangement of the segmented capillaries used for free molecular flow simulations (using Molflow+).
	The dimensions are same as that used in experiments.
	The texture plot on the top and bottom show the spatial distribution of He atoms (obtained using simulations) emerging from the segmented and a single long capillary of the same overall dimensions, respectively.
	(b) Comparison of angular distributions obtained from simulations and experiments. 
	Observed width (FWHM) of angular distributions is 0.7$^\circ$ and 0.8$^\circ$ from experiments and simulations, respectively.
	The position 0$^\circ$ corresponds to the centerline of the beam.}
	\label{fig:expt_molflow_comp}
\end{figure*}

Sticking probability measurements of oxygen on Cu(110) surface, using He reflectivity, were performed in an independent ultrahigh vacuum (UHV) chamber.
This custom designed UHV chamber is a part of a recently built experimental setup for quantum state resolved molecule-surface scattering and reactivity measurements (Fig. \ref{fig:expt_setup_panel}b).
A Cu(110) single crystal (99.9999\% pure, 10 mm diameter and 2 mm thickness), cut to precision better than 0.1$^\circ$ and polished to have roughness lower than 10 nm (MaTeck Material Technologie \& Kristalle GmbH) was used as a target sample.
It was mounted on a 4-axis differentially pumped manipulator (XYZ$\mathrm{\Theta}$) using a pair of 0.25 mm diameter tungsten wires which enabled sample heating.
The sample manipulator is equipped with electrical and thermocouple feedthroughs for resistive heating and monitoring sample temperature (using a K-type thermocouple).
This UHV chamber is also equipped with Ar ion source (IS40, Prevac) for surface cleaning via sputtering and Auger electron spectrometer (AES, Model: SMG600, OCI Vacuum Microengineering) for checking surface chemical composition.
Additional vacuum chambers denoted by F, G and H in Fig. \ref{fig:expt_setup_panel}b correspond to a double differentially pumped system, built for quantum state selected molecular beam - surface scattering experiments.
These stages were not used in the present work and were isolated from the UHV chamber using a custom-built Teflon sealed sliding valve.

Initial surface cleaning was done using repeated sputtering and annealing cycles, similar to that reported  previously \cite{musket_surface_cleaning_rev_1982}.
Thereafter, for day-to-day operation, the sample surface was subjected to Ar ion sputtering for a duration of 30 minutes (0.4 $\mu$A ion current) at 3 keV ion energy.
Under these conditions, impurity levels (mainly carbon) were found to be below the detection threshold of AES.
Subsequently, the surface was annealed at 800 K for 20-30 min and allowed to cool down to 300-310 K before conducting the He reflectivity measurements. 
Base pressure of the UHV chamber in these measurements was $1.5\times 10^{-9}$ mbar. 
Under these conditions, we observed that the sample remained clean (as measured by AES) for a duration of 4 hours, which is sufficient for the sticking probability measurements under consideration (typically 15 minutes per measurement).

Source chamber with the segmented capillary was mounted on one of the arms of UHV chamber. 
Alignment of the capillary with the target surface was checked by sending a laser beam through the capillary inlet.
Appropriate sample position was determined by ensuring that the light beam exiting the capillary lands on the center of the sample surface and the reflected light beam enters the sampling aperture. 
Thereafter, the chamber was pumped and baked out to reach UHV conditions needed for sticking probability measurements.
A well collimated He beam emerging from capillary outlet, denoted by A in Fig. \ref{fig:expt_setup_panel}b, was made incident on a clean Cu(110) surface at an incidence angle of 50$^{\circ}$ from the surface normal.
Specularly reflected He atoms entered a differentially pumped sampling tube through a slit (1 mm width, 20 mm length) and were detected using a mass spectrometer (SRS RGA 200), in a manner similar to that used for beam width characterization. 
The UHV chamber was pumped by a turbomolecular pump (HiPace 700 H, Pfeiffer Vacuum) which was backed by a dry roots pump (ACP 15, Pfeiffer Vacuum).
He source and the detection stage were pumped by independent turbomolecular pumps (HiPace 80, Pfeiffer Vacuum). 
These were backed by a rotary vane pump (Duo 11, Pfeiffer). 
Steady state pressures in the source and the UHV chamber were 2$\times$10$^{-8}$ mbar and 1.5$\times$10$^{-9}$ mbar with He beam off.  
With the He beam on, these values were 2$\times$10$^{-6}$ mbar and 2$\times$10$^{-9}$ mbar (Cu(110) surface moved away), respectively.
Under these conditions, background partial pressure of He in the sampling tube was 5$\times$10$^{-11}$ mbar (He beam off) and increased to 8$\times$10$^{-11}$ with He beam on. 
With the Cu(110) surface in optimal position, specular reflected signal for He from a clean Cu(110) surface typically corresponded to 2$\times$10$^{-10}$ mbar, as seen by the mass spectrometer.

For the surface coverage dependent He reflectivity and sticking probability measurements, high purity oxygen gas was leaked into the chamber using a precision leak valve (simultaneously with the He beam on).
Steady state background pressure with the oxygen leaking in was set to 1$\times$10$^{-8}$ mbar, corresponding to an incident oxygen flux of approximately 0.01 ML per second.
Time integrated oxygen pressure was measured by an ionization gauge and was used to evaluate the incident oxygen dose on the sample.
Change in reflected He signal, normalized to the reflectivity of the clean surface (I/I$_0$) and the incident oxygen dose were used to obtain the sticking probabilities and diffuse elastic scattering cross section. 

\section{Results and discussions}

\subsection{\label{sec:level2} Segmented capillary source characterization}

Figure \ref{fig:expt_molflow_comp}a shows a snapshot of free molecular flow simulations (using Molflow+) of He flowing through a segmented capillary structure with dimensions similar to that used in our measurements.
The two dimensional texture plot (top) depicts spatial distribution of the generated He beam, mapped at a distance of 160 mm (same as in beam width measurements) from the exit plane.
Texture plot shown below depicts the distribution resulting from a single capillary with the same dimensions (diameter = 0.5 mm and length = 50 mm, no segmented structure).
Quite clearly, the segmented structure leads to a much narrower and well collimated beam.
Figure \ref{fig:expt_molflow_comp}b depicts a comparison among the angular distributions of output flux obtained from the experiment and simulations.
It is quite evident that the experimental observations are reproduced well by these simulations.
Most importantly, a narrow beam with an angular width of  0.7$^\circ$ (full width half maximum, FWHM), corresponding to a beam size of 1.9 mm at a distance of 160 mm is observed in our measurements.
Considering a source size of 0.5 mm and that the sampling slit is 1 mm wide, we estimate the angular divergence of the He beam to be 7 mrad (FWHM).
Based on the pressure changes and the observed beam width we estimate the flux of He atoms in the beam to be $\sim 10^{18}$ atoms/(sec str).
Differential pumping stages enabled by the open segments effectively suppress the broad tail-like feature in the angular distribution, expected for long thin capillaries \cite{scoles_book}. 
Such a collimated beam with is well-suited for He atom reflectivity measurements providing a high signal to background ratio.

\subsection{\label{sec:level2} Oxidation of Cu(110) monitored using He reflectivity}

\begin{figure}
	\centering
	\includegraphics[width = 1\linewidth,trim={0.2cm 0.3cm 1.9cm 1.7cm},clip, draft=false]{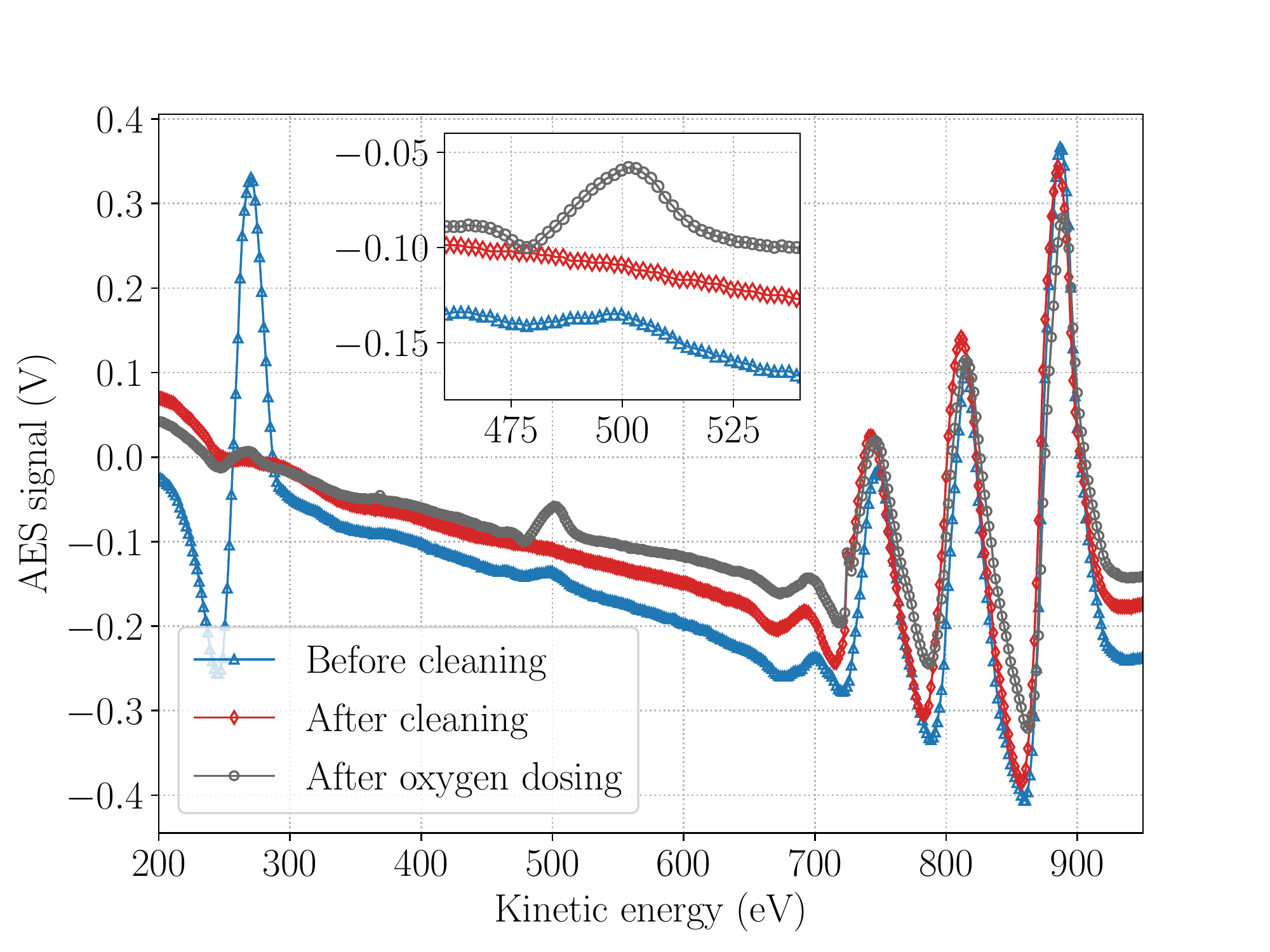}
	\caption{
		Auger electron spectra of the Cu(110) surface, measured before cleaning (lower curve, blue), after cleaning (middle curve, red) and after the oxygen dosing (approximately 25 ML). Characteristic peaks at 272 eV and 503 eV correspond to carbon and oxygen on the sample while peaks in the 700 - 920 eV region correspond to Cu. 
		Inset shows a zoomed view of the peaks resulting from oxygen.}
	\label{fig:auger_spectra}
\end{figure}

\begin{figure*}[!]
	\centering
	\includegraphics[width = 0.9\linewidth, draft=false]{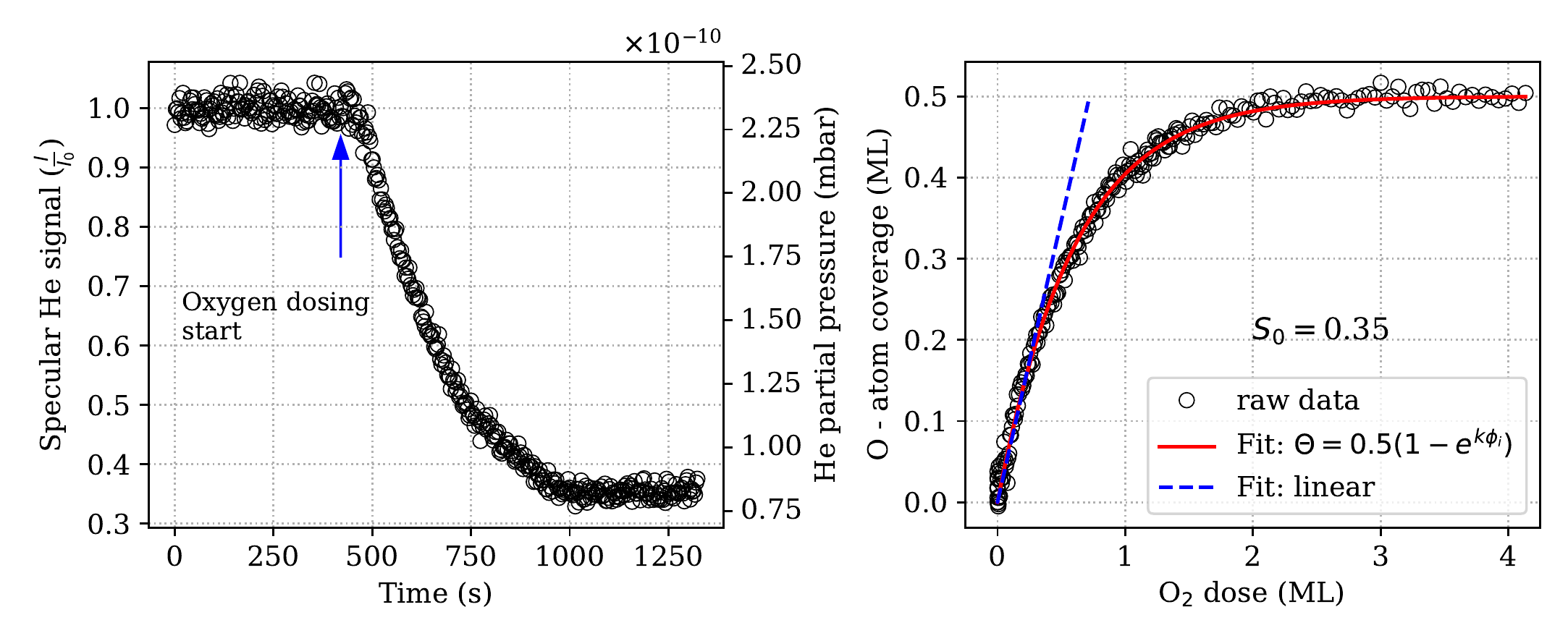}
	\caption{(left) Specular He signal (normalized) of the Cu(110) surface as a function of time, as the surface was exposed to oxygen (red arrow).
	The reflected He signal as observed by the mass spectrometer is also shown on the y-axis on the right  
	(b) Surface coverage of O-atoms as a function of oxygen exposure (in ML), obtained using the data shown in the left panel.
	Red curve shows a fit to a model assuming 0.5 ML as the saturation coverage.
	Blue line (dashed) shows a linear fit to the initial part of the curve. We estimate the initial sticking probability to be 0.35 (using linear fit).}
	\label{fig:He_ref}
\end{figure*}

Surface cleanliness of the Cu(110) sample was checked using AES (Fig. \ref{fig:auger_spectra}). 
The lower (blue) and the middle (red) curves represent the spectrum obtained before and after sample cleaning (Ar ion sputtering), respectively.
Once the major contaminants carbon (272 eV) and oxygen (503 eV) were removed and only prominent features corresponding to copper (700-920 eV) remained, the sample was annealed (800 K, 30 min) and cooled down to 300 K.
Following this, the clean Cu(110) surface was exposed to oxygen, leaked into the chamber through a precision leak valve, corresponding to a dose of approximately 25 ML.
Subsequent AES measurements (Fig. \ref{fig:auger_spectra}, grey curve) showed that the surface is covered with oxygen.
It was also noted that under these conditions the surface oxygen coverage was saturated and no further increase in oxygen signal was observed with additional oxygen exposure.
This is consistent with previous studies where dissociative chemisorption of oxygen on Cu(110) has been studied and on exposures greater than 10 ML the surface coverage was observed to be saturated \cite{ertl_Cu_O2_LEED_German_1967, gruzalski_Cu_O2_reconstruction_1984, gruzalski_Cu_O2_xps_1985, jensen_Cu_O2_reconstruction_STM_1990}

Having established appropriate conditions for preparing the oxygenated Cu(110) surface in a controlled manner, we measured the evolution of surface coverage in real-time using specular He reflection as a probe.
Upon exposure to oxygen, normalized reflected He signal (I/I$_0$)  decreased with time and ultimately reached a steady state value of about 0.4 times its initial value (Fig. \ref{fig:He_ref}a).
These observations clearly show that as the oxygen coverage on the Cu(110) surface builds up, He reflectivity decreases.
At longer times ($>$ 1000 sec), the surface was saturated with oxygen (as confirmed by AES) and no further change in He reflectivity was observed.
This decrease in specular reflected He signal is attributed to increased diffuse scattering of incident He atoms on the oxygen covered Cu(110) surface.
These observations are consistent with those expected from several He scattering studies reported previously \cite{comsa_HeScat_rev_1985, farias_rev1998}, where the adsorbate induced disorder on the surface leads to reduced specular reflection of He atoms.

In order to evaluate surface coverage and initial sticking probability using this data, a relation among He reflectivity  and surface coverage needs to be established first.
Previous studies using low energy electron diffraction show that saturation of oxygen on Cu(110) surface corresponds to a coverage of 0.5 ML.
Using this information and the fact that steady state He reflected signal corresponds to an oxygen saturated surface (as confirmed using AES), we obtain the following relation: 
I/I$_0$ = 1 corresponds to zero coverage and the steady state  He reflectivity ($\sim$0.4) following oxygen exposure corresponds to 0.5 ML oxygen coverage (saturation).
Figure \ref{fig:He_ref}b shows a plot of surface coverage vs oxygen dose obtained using the above method.
Oxygen dose in terms of monolayers was estimated from the change in background pressure (after leaking in oxygen) and considering surface atom density of Cu(110) to be 1.09$\times 10^{15}$ atoms/cm$^2$.
Quite clearly, the observed trend follows the expected behaviour where the rate of sticking is proportional to the number of unoccupied adsorption sites available.
Red line (Fig. \ref{fig:He_ref}b) corresponds to a best fit using a model 0.5(1-e$^{k \phi_{\rm i}}$), where $\phi_{\rm i}$ represents the incident O$_2$ dose and the initial sticking probability, $S_0 = k/4$ (considering that saturation coverage is 0.5 ML and each O$_2$ dissociation gives rise to two O-atoms adsorbed on the surface).
The dashed blue curve represents a linear fit to the initial part of the curve which results in an initial sticking probability of 0.35 (= slope/2).

It should be noted that for the same system, a sticking 
probability of 0.23 has been reported previously \cite{gruzalski_Cu_O2_xps_1985, pudney_Cu_O2_ActivatedDiss_1990, nesbitt_Cu_O2_Chemisorption_1991}. 
These systematic differences are likely to be arising from the fact that in our case oxygen dosage estimation was made using the pressure values directly obtained from the ion gauge using the typical gas sensitivity factors, without any additional calibration.
Nonetheless, we checked the repeatability of our observations by performing an additional five independent measurements using the same method (see SI-1). 
Overall, the $S_0$ values obtained ranged from 0.33 to 0.37, showing good consistency among the results and validate the utility of this method. 
Additionally, based on the initial rate of change in I/I$_0$ with respect to O-atom coverage, we estimate the diffuse elastic scattering cross section ($ = \frac{d(I/I_0)}{d\Theta}$) \cite{comsa_diffuse_1983} of He from the adsorbed O-atoms to be 90 $\rm {\AA^2}$ (assuming $S_0$ = 0.23).

\section{Concluding Remarks}
In this work, we have successfully demonstrated the design, development and characterization of a simple, compact atomic beam source based on a segmented capillary design.  
It produces a highly collimated beam of He atoms with angular divergence of 7 mrad and a brightness of $\sim$ 10$^{18}$ atoms /sec /str.
Further, we demonstrate an application of this compact atomic beam source by using it for measuring the real time surface coverage and initial sticking probability of oxygen on a  clean Cu(110) surface, by means of measuring its He reflectivity.
The compact footprint and relatively simpler design, unlike conventional differentially pumped vacuum chamber systems, allows for a relatively easier integration into our molecule - surface scattering experimental setup.
This design is flexible in the sense that, the angular width can be easily adjusted by choosing an appropriate L/d ratio and segment length, without having to make any major changes to the vacuum chamber itself.

We believe that this design is very valuable for quantum state resolved chemisorption experiments, currently being developed in our lab. 
Here, He reflectivity measurements using such a compact and simple source allows for a very practical way of measuring surface coverage in real time, non-destructive, highly sensitive and universal manner.
This design can be further improved by using an optical window at the inlet side of the capillary.
This will allow the in-vacuum alignment (using a laser beam) of the atomic beam, which is currently not possible in the present setup. 
Further, using two sequential slits for additional differential pumping in the detection setup is expected to provide a much higher background rejection, leading to a higher signal to background ratio and ultimately better detection sensitivity. 
These improvements will be considered for future versions of this set up. 
Additionally, we also envisage that such a design is potentially useful for real-time, non destructive monitoring of the growth of thin films on atomically flat surfaces, especially where layer by layer growth occurs. 
Also, such a design is expected to be generally useful in several situations where a highly directed flux of atoms/molecules is required with a compact footprint.

\section*{Supplementary Information}

\begin{itemize}
	\item SI-1: Repeated surface coverage measurements
\end{itemize}

\section*{Data Availability}
All relevant data related to the current study are available from the corresponding author upon reasonable request.

\section*{Acknowledgements}
This work was partly supported by intramural funds at TIFR Hyderabad from the Department of Atomic Energy and Scientific and Engineering Research Board, Department of Science and Technology (grant numbers: ECR/2018/001127 and CRG/2020/003877). 
We thank Rakesh Moodike (institute workshop) for suggesting the use of stainless steel capillary, fabricating the segmented structure and components for the detection assembly.

\section*{Author Contributions}
GB performed the simulations, designed the necessary components and characterized the performance of the segmented capillary source with inputs from PRS. 
SKS contributed to preparing the UHV chamber used to conduct the sticking probability measurements with inputs from PRS. 
GB and SKS performed the sticking probability measurements using He reflectivity and analyzed the data.
PRS conceptualized the project. 
GB and PRS prepared the manuscript with inputs from SKS.
All authors discussed the results, analysis and contributed to the manuscript.

\bibliographystyle{unsrt}
\bibliography{references}

\end{document}